\documentclass[cpp,a4paper,fleqn]{w-art}

\usepackage{times,cite,w-thm}
\theoremstyle{plain}

\theoremstyle{definition}

\usepackage[]{graphicx}

\usepackage{subfigure}
\usepackage[english]{babel}

\begin{document}

\DOIsuffix{theDOIsuffix}

\Volume{XX}
\Month{XX}
\Year{XXXX}

\pagespan{1}{}

\Receiveddate{XXXX}
\Reviseddate{XXXX}
\Accepteddate{XXXX}
\Dateposted{XXXX}

\subjclass[pacs]{12.38Mh, 31.15.Qg, 51.20.+d, 52.27Gr}
\keywords{strongly correlated plasma, quark gluon plasma}

\title[Nonideal quark-gluon plasma]{Equation of state of strongly coupled quark--gluon plasma --\\
Path integral Monte Carlo results}
\author[V. Filinov]{V.S.~Filinov\inst{1}
  \footnote{Corresponding author\quad E-mail:~\textsf{filinov@mail.ru},
            Phone: +\,49\,431\,880-4063,
            Fax: +\,49\,431\,880-4094}}
\author[M. Bonitz]{M. Bonitz\inst{2}}
\author[Y.B.~Ivanov]{Y.B.~Ivanov\inst{3,4}}
\author[V.V.~Skokov]{V.V.~Skokov\inst{3,5}}
\author[P.R. Levashov]{P.R. Levashov\inst{1}}
\author[V.E. Fortov]{V.E. Fortov\inst{1}}
\address[\inst{1}]{Joint Institute for High Temperatures, Russian Academy of
Sciences, Izhorskaya 13/19, Moscow 127412, Russia}
\address[\inst{2}]{Institute for Theoretical Physics and Astrophysics, Christian Albrechts University Kiel, \\ Leibnizstrasse 15, D-24098 Kiel, Germany}
\address[\inst{3}]{Gesellschaft fur Schwerionenforschung, Planckstrasse 1, D-64291 Darmstadt, Germany}
\address[\inst{4}]{Russian Research Center ``Kurchatov Institute'', Kurchatov Sq. 1,  123182 Moscow, Russia}
\address[\inst{5}]{Bogoliubov Lab. of Theoretical Physics, Joint Institute for Nuclear Research, 141980, Dubna, Russia}

%---------------------------

\date{\today}% It is always \today, today,

\begin{abstract}
A strongly coupled plasma of quark and gluon quasiparticles at temperatures from $ 1.1 T_c$ to $3 T_c$ is studied by path integral Monte Carlo simulations. This method extends previous classical nonrelativistic simulations based on a color Coulomb interaction to the quantum regime. We present the equation of state and find good agreement with lattice results. Further, pair distribution functions and color correlation functions are computed indicating strong correlations and liquid-like behavior.
\end{abstract}

\maketitle

\section{Introduction}\label{s:intro}
Determining the properties of deconfined quark-gluon plasma (QGP) is one of the main challenges of strong-interaction physics both theoretical and experimental. Many features of this matter were experimentally discovered at the Relativistic Heavy Ion Collider (RHIC) at Brookhaven. The most striking result, obtained from analysis of these experimental data \cite{shuryak08}, is that the deconfined quark-gluon matter behaves as almost perfect fluid rather than as a perfect gas, as it could be expected from the asymptotic freedom.

From the theory side, the most fundamental way to compute properties of strongly interacting matter is provided by lattice QCD, see the recent review \cite{Lattice09}. Interpretation of these computations requires application
of various QCD motivated, albeit schematic, models simulating various aspects of the full theory and allowing for a deeper physical understanding.
The above mentioned strongly correlated behavior of the QGP is expected to show up in long-ranged spatial correlations of quarks and gluons which, in fact, may give rise to liquid-like and, possibly, solid-like structures. This expectation is based on very similar behavior observed in electrodynamic plasmas, as was pointed out e.g. in Refs. \cite{thoma04,shuryak1}. This similarity has been exploited to formulate classical non-relativistic models of a color Coulomb interacting QGP \cite{shuryak1} which is numerically analyzed by classical molecular dynamics simulations. While this has allowed to incorporate nonideality effects, quantum effects were either neglected or included phenomenologically via a short-range repulsive correction to the pair potential, e.g. \cite{shuryak1}. However, such a rough model may become a critical issue at higher densities where quantum and Fermi statistics effects of the quarks should have a strong influence on the properties of the QGP. Similar models had been used in electrodynamic plasmas and showed poor behavior in the region of strong wave function overlap, in particular at the Mott density.

This difficulty can be eliminated by deriving effective quantum potentials, as was shown by some of the present authors before \cite{afilinov_jpa03,afilinov_pre04,ebeling06}. Following an idea of Kelbg \cite{kelbg} quantum corrections to the pair potential can be rigorously derived in perturbation theory with respect to the coupling parameter \cite{dusling09}. To extend the method of quantum potentials to larger coupling an ``improved Kelbg potential'' was derived which contains a single free parameter
which is fitted to the exact solution of the quantum-mechanical two-body problem and exactly reproduces the thermodynamic properties up to moderate couplings \cite{afilinov_pre04}. However, this approach may fail if the system forms bound states of more than two particles leading to a break-down of the pair approximation for the density matrix, as was demonstrated in Ref. \cite{afilinov_pre04}. A superior approach which does not have these limitations, is to use the original Kelbg potential in path integral Monte Carlo (PIMC) simulations which effectively map the problem onto a high-temperature weakly coupled and weakly degenerate one. This allows one to rigorously extend the analysis to strong couplings and is, therefore, the method of choice for the present purpose. Beside the nonideality and quantum effects our approach  takes into account the effects of the Fermi (Bose) statistics of quarks (gluons) by a proper antisymmetrization (symmetrization) of the $N-$body density matrix. For temperature and density of the QGP considered in this paper which are similar to the conditions of Ref. \cite{shuryak1} these effects are very important as the quasiparticle thermal wave length is larger than the average interparticle distance. This is, in particular, important for the behavior of the pair distribution functions (see below).

Here, we develop a PIMC approach to the strongly coupled QGP which takes the Fermi (Bose) statistics of quarks (gluons) and quantum degeneracy selfconsistently into account. This method has been successfully applied to strongly coupled electrodynamic plasmas before, e.g. \cite{filinov_ppcf01,bonitz_jpa03,bonitz_pop08}. Examples are partially ionized dense hydrogen plasmas where liquid-like and crystalline behavior was observed \cite{filinov_jetpl00,bonitz_prl05}. Moreover, also partial ionization effects and pressure ionization could be studied from first principles \cite{filinov_jpa03}. The same methods have been applied also to electron-hole plasmas in semiconductors \cite{bonitz_jpa06,filinov_pre07}, including excitonic bound states, which have many similarities to the QGP due to the smaller mass differences as compared to electron-ion plasmas.

In this paper we present first exploratory PIMC simulations of a nonideal quark-gluon plasma.
The main goal is to test this approach for ability to reproduce the equation of state known from lattice data \cite{Lattice09}. To this end we use the simplest model of a QGP consisting of quarks, antiquarks and gluons interacting via a color Coulomb potential due to Gelman et al. \cite{shuryak1} with several approximations for the temperature dependence of the quasiparticle masses. We report surprisingly good agreement with the lattice data for one of the parameter sets, which gives us confidence that the model correctly captures main properties of the nonideal QGP.

The paper is organized as follows. In Sec.~\ref{s:model} we introduce the model and approximations used which is followed by an overview on our PIMC simulations in Sec.~\ref{s:pimc}. Sec.~\ref{s:results} contains our results on the equation of state and on the various pair distribution functions of the QGP, and we conclude in Sec.~\ref{s:discussion} with a discussion of the results together with an outlook on further improvements of the approach.

%----------------------------
\section{Theoretical Model}\label{s:model}

Our model is based on precisely the same assumptions as those in Ref. \cite{shuryak1}
which are summarized as follows: 
\begin{description}
 \item[I] All particles (quarks and gluons) are heavy, i.e., $m > T$, where $m$ is the mass of a particle and $T$ the temperature and, therefore, they move non-relativistically.
This assumption is based on the analysis of lattice data \cite{Lattice02,LiaoShuryak}.
 \item[II] Since the order of magnitude of quark and gluon masses, deduced from the
lattice data \cite{Lattice02,LiaoShuryak} is the same, we do not distinguish these masses and put them equal. Moreover, because of the latter we do not distinguish between quark flavors. 
 \item[III] The interparticle interaction is dominated by a color-electric Coulomb interaction, see Eq. (\ref{Coulomb}). Magnetic effects are neglected as sub-leading ones, in the nonrelavistic limit.
 \item[IV] The color operators $t^a$ are substituted by their average values, i.e. by classical color vectors,
relying on the fact that the color representations are large.
\end{description}
The quality of these approximations and their limitations were discussed in Ref. \cite{shuryak1}.

Aiming at a first test of this model in PIMC simulations, in this paper, we are going to consider the QGP only at zero baryon density. Therefore, this model requires the following quantities as an input:
\begin{enumerate}
 \item the temperature dependence of the quasiparticle mass, $m(T)$,
 \item the density of particles, $n(T)$, at a given temperature,  following \cite{shuryak1},
where we assume that the numbers of quarks, antiquarks and gluons are approximately equal,
 \item the coupling constant, $g^2(T)$, at a given temperature, see Eq. (\ref{Coulomb}).
[Note that, because of the running coupling in the QCD, $g^2$ generally depends on $T$].
\end{enumerate}
All the input quantities should be deduced from the lattice data or from an appropriate model simulating these data.

%----------------------------
\section{Path integral Monte Carlo Simulations}\label{s:pimc}
As discussed in section \ref{s:model} we consider a three-component QGP consisting of
$N=N_q+N_{\bar{q}} +N_g$
quasiparticles, where $N_q, N_{\bar{q}}, N_g$ are, respectively, the number of (dressed)  quarks, antiquarks and   gluons  in thermal equilibrium, so the  temperature-dependent Hamiltonian can be written as
${\hat{H}_\beta}={\hat{K}_\beta}+{\hat{U}_\beta}^c$, where $\beta=1/k_BT$, is the inverse temperature and $k_B$ is Boltzmann's constant. Here we introduced the kinetic and color Coulomb interaction energy of the quasiparticles
\begin{eqnarray}
\label{Coulomb}
{\hat{K}_\beta}=\sum_{t=1}^N
\frac{p^2_t}{2m_t(\beta)},
\qquad
{\hat{U}_\beta}^c=\frac{1}{2}\sum_{p=1}^N\sum_{t=1}^N \frac{C_{pt}g^2
\langle Q_p|Q_t \rangle}{4\pi|r_p-r_t|},
\end{eqnarray}
Here the $Q_p$ denote Wong's color variables which are $3D$ unit vectors, and the constants $C_{pt}=C_{tp}$ are products of eigenvalues of the Casimir operator \cite{shuryak1}:
% We have tested  two cases. In the first one 
$C_{qq}=C_{\bar{q}q}=C_{\bar{q}\bar{q}}=4/3$,  
$C_{qg}=C_{\bar{q}g}=2$ and $C_{gg}=3$ with $g^2 = 2\pi$. 

The thermodynamic properties in the
canonical ensemble with given temperature $T$ and fixed volume $V$ are fully described by the density operator ${\hat \rho} = e^{-\beta {\hat H}}/Z$ with the partition function (normalization constant)
\begin{equation}\label{q-def}
Z(N_q,N_{ \bar{q}},N_g,V;\beta) = \frac{1}{N_q!N_{ \bar{q}}!N_g!} \sum_{\sigma}\int\limits_V
dr dQ\,\rho(r,Q, \sigma ;\beta),
\end{equation}
where $\rho(r,Q, \sigma ;\beta)$ denotes the diagonal matrix
elements of the density operator at a given value $\sigma$ of the total spin. In Eq.~(\ref{q-def}), $r=\{r_q,r_{ \bar{q}},r_g\}$ and
$Q=\{Q_q,Q_{ \bar{q}},Q_g\}$ are the spatial and color coordinates,
while $\sigma=\{\sigma_q,\sigma_{ \bar{q}}\}$
are the spin degrees of freedom% of quarks, antiquarks and gluons
, i.e.
$r_a=\{r_{1,a}\ldots r_{l,a}\ldots r_{N_a,a}\}$, $Q_a=\{Q_{1,a}\ldots Q_{l,a}\ldots Q_{N_a,a}\}$ and $\sigma_a=\{\sigma_{1,a}\ldots \sigma_{l,a}\ldots
\sigma_{N_a,a}\}$ with $a=q, \bar{q},g$.

In order to calculate thermodynamic functions, the logarithm of
the partition function has to be differentiated with respect to
thermodynamic variables. For example, for pressure and internal
energy follows
\begin{eqnarray}
\beta P &=& \partial {\rm ln} Z / \partial V = [\alpha/3V \partial
{\rm ln} Z / \partial \alpha]_{\alpha=1}, \label{p_gen}
\\
\beta E &=& -\beta \partial {\rm ln} Z / \partial \beta,
\label{e_gen}
\end{eqnarray}
where $\alpha= L/L_0$ is a length scaling parameter.

Of course, the exact density matrix of interacting quantum
systems is not known (particularly for low temperatures and high
densities), but it can be constructed using a path integral
approach %~\cite{feynman-hibbs}
based on the operator identity
$e^{-\beta {\hat H}}= e^{-\Delta \beta {\hat H}}\cdot
e^{-\Delta \beta {\hat H}}\dots  e^{-\Delta \beta {\hat H}}$,
where the r.h.s. contains $n+1$ identical factors with $\Delta \beta = \beta/(n+1)$, which allows us to
rewrite the integral in Eq.~(\ref{q-def})
\begin{eqnarray}
&&\sum_{\sigma} \int\limits dr^{(0)}dQ^{(0)}\,
\rho(q^{(0)},Q^0,\sigma;\beta) =
%\nonumber\\&&
\int\limits  dr^{(0)}dQ^{(0)} \dots
dr^{(n)}dQ^{(n)} \, \rho^{(1)}\cdot\rho^{(2)} \, \dots \rho^{(n)}
\times
\nonumber\\
\nonumber\\&&
\sum_{\sigma}\sum_{P_q} \sum_{P_{ \bar{q}}}\sum_{P_g}(- 1)^{\kappa_{P_q}+ \kappa_{P_q}} \,
{\cal S}(\sigma, {\hat P_q}{\hat P_{ \bar{q}}}{\hat P_g} \sigma^\prime)\,
%\times \nonumber\\&&
{\hat P_q} {\hat P_{ \bar{q}}}{\hat P_g}\rho^{(n+1)}\big|_{r^{(n+1)}= r^{(0)}, \sigma'=\sigma}\,.
 \label{rho-pimc}
\end{eqnarray}
The spin gives rise to the spin part of the density matrix (${\cal
S}$) with exchange effects accounted for by the permutation
operators  $\hat P_q$, $\hat P_{ \bar{q}}$ and $\hat P_g$ acting on the quark, antiquark and gluon spatial $r^{(n+1)}$ and color $Q^{(n+1)}$ coordinates and spin projections $\sigma'$. The
sum is over all permutations with parity $\kappa_{P_q}$ and
$\kappa_{P_{ \bar{q}}}$. In Eq.~(\ref{rho-pimc}) the index $k=1\dots n+1$
labels the off-diagonal high-temperature density matrices
$\rho^{(k)}\equiv \rho\left(r^{(k-1)}Q^{(k-1)},r^{(k)}Q^{(k)};\Delta\beta\right) =
\langle r^{(k-1)}|e^{-\Delta \beta {\hat H}}|r^{(k)}\rangle\delta(Q^{(k-1)}-Q^{(k)})$.
Accordingly each particle is represented by a set of $n+1$ coordinates
(``beads''), i.e. the
whole configuration of the particles is represented by a
$3(N_q+N_{ \bar{q}}+N_g)(n+1)$-dimensional vector
$\tilde{r}\equiv\{r_{1,q}^{(0)}, \dots r_{1,q}^{(n+1)},
r_{2,q}^{(0)}\ldots r_{2,q}^{(n+1)}, \ldots r_{N_q,q}^{(n+1)};
r_{1, \bar{q}}^{(0)}\ldots r_{N_g,g}^{(n+1)} \}$ and a $2(N_q+N_{ \bar{q}}+N_g)$-dimensional color vector $\tilde{Q}\equiv\{Q_{1,q}^{(0)},Q_{2,q}^{(0)},
\ldots, Q_{N_g,g}^{(0)}\}$.
The main contributions to the partition function come from
configurations in which the `size' of the cloud of beads of quasiparticles is of
the order of their thermal  wavelength, whereas
typical distances  between beads of each quasiparticle are of the order of the
wavelength taken at the $(n+1)$-times higher temperature $\Delta \beta$.
To determine the pressure or total energy  in the path integral
representation~(\ref{rho-pimc}) each high-temperature density matrix has to be
differentiated in turn according to expressions (\ref{p_gen}) and (\ref{e_gen}).

Let us now consider approximations for the high-temperature density matrices $\rho_{sk}$.
An approximation which is suitable for direct PIMC simulations has the following form, generalizing the 
electrodynamic plasma results \cite{filinov_ppcf01} to the case of an additional bosonic species (the gluons):
\begin{eqnarray}
\rho_{sk}(r,\beta) &=& C^s_{N_q} C^k_{N_{ \bar{q}}} \, e^{-\beta
U(r,Q,\beta)} \prod\limits_{l=1}^n \prod\limits_{p=1}^N
\phi^l_{pp}
%\nonumber\\&&
{\rm det}\,||\tilde{\phi}^{n,1}||_{sk}\times{\rm per}\,||\tilde{{\phi}}^{n,1}||, \label{rho_s}
\end{eqnarray}
where $s(k)$ is the number of quarks (antiquarks) with the same spin projection, antisymmetrization and  symmetrization are taken into account by the symbols ``det'' and ``per'' denoting the determinant and permanent, respectively. Further, we introduced the total color interaction energy
\begin{eqnarray}
U(r,Q,\beta) =
\sum_{l=0}^{n}
U\left(r^{(l)},Q,\Delta\beta\right)/(n+1), 
\label{up}
\end{eqnarray}
Here, the result is rewritten in terms of dimensionless coordinates $r$ which depend on the dimensionless distances between neighboring beads, $\xi_{1}^{(1)}, \dots \xi_{N}^{(n)}$, according to $r\equiv [r^{(1)}; r^{(1)}+y^{(1)};r^{(2)}+y^{(2)}; \dots;r^{(n)}+y^{(n)}]$, with $y^n=\Delta\lambda\sum_{k=1}^{n}\xi^{(k)}$. Further, we introduced the high-temperature
De Broglie wavelength, $\Delta\lambda_a^2=2\pi\hbar^2 \Delta\beta/m_a(\beta)$, and the exchange matrix
$\phi^l_{pp}\equiv \exp\left[-\pi\left|\xi^{(l)}_p\right|^2\right]$.

The path integral representation of the density matrix is exact in the limit $n\to \infty$. For any finite 
number $n$, the error of the above approximations for the whole product on the r.h.s. of Eq.
(\ref{rho-pimc}) is of the order $1/(n+1)$ whereas the error of each high-temperature factor is
of the order $1/(n+1)^2$, as was shown in Ref. \cite{filinov_ppcf01}. Our
approximation of the high-temperature density matrix is given by products of two-particle density matrices
$\rho^{(i)}=\rho_0^{(i)}\rho_U^{(i)}+O[(1/n+1)^2]$, where
$\rho_0^{(i)}$ is the kinetic density matrix, while
$\rho_U^{(i)}=e^{-\Delta \beta U(r^{(i-1)},Q)}\delta(r^{(i-1)}-r^{(i)})$, where $U$ denotes the pair sums of the off-diagonal two-particle effective quantum potentials.
These potentials are straightforward generalizations of the corresponding potentials of electrodynamic plasmas  \cite{filinov_ppcf01,afilinov_pre04} to the case of color Coulomb interaction,
$\Phi^{pt}({\bf r}_p,{\bf r}_p', {\bf r}_t, {\bf r}_t', Q)$. % \cite{filinov76}.
In the following we will use the diagonal element (${\bf r}'_{p}={\bf r}_{p},
{\bf r}'_{t}={\bf r}_{t}$) which will be called ``color
Kelbg potential'' and depends only on a single distance.
It is derived by approximating the off-diagonal matrix elements of the effective binary interaction
 by the diagonal ones at the center coordinate
$\Phi^{ab}(r,r',Q_a,Q_b;\Delta\beta)\approx
\Phi^{ab}(\frac{r+r'}{2},,Q_a,Q_b;\Delta\beta)$ or
$\Phi^{ab}(r,r';,Q_a,Q_b, \Delta\beta)\approx [\Phi^{ab}(r,,Q_a,Q_b,\Delta\beta) + 
\Phi^{ab}(r',Q_a,Q_b, \Delta\beta)]/2$.
The result for the diagonal color Kelbg potential is (we retain the same notation as before)
\begin{eqnarray}
&&\Phi^{pt}({\bf r}_{p},{\bf r}_{p}, {\bf r}_{t}, {\bf r}_{t},Q_p,Q_t,\Delta\beta) =
\Phi^{pt}(|{\bf r}_{pt}|,Q_p,Q_t,\Delta\beta)
\, \nonumber\\
&=& \frac{C_{pt} \, g^2\,\langle Q_p|Q_t \rangle}{4 \pi \lambda_{pt} x_{pt}} \,\left[1-e^{-x_{pt}^2} +
\sqrt{\pi} x_{pt} \left(1-{\rm erf}(x_{pt})\right) \right],
\label{kelbg-d}
\end{eqnarray}
where $x_{pt}=|{\bf r}_{p}-{\bf r}_{t}|/\lambda_{pt}$. Note that the color Kelbg potential approaches the color Coulomb potential at distances larger than the De Broglie wavelength. Most importantly, it is finite at zero distance (it is of the order of $T$), removing in a natural way the classical divergences which makes any artificial cut-offs obsolete.

Finally let us comment on the treatment of the exchange properties of quarks, antiquarks and gluons.
The density matrix (\ref{rho_s}) has been transformed to a form which does not
contain an explicit sum over permutations and thus no sum of
terms with alternating sign (in the case of quarks and antiquarks). Instead, the whole exchange problem
is contained in exchange matrices from which we have to compute the determinant (for quarks and antiquarks) or the permanent (for gluons),
\begin{eqnarray}
||\tilde{\phi}^{n,1}_{pt,ol}||_{sk}\equiv
\left|\left|\phi^{n,1}_{pt}\cdot \frac{\langle Q_p|Q_t\rangle+1}{2}\right|\right|_{s} \cdot
\left|\left|\phi^{n,1}_{ol}\cdot \frac{\langle Q_o|Q_l\rangle+1}{2}\right|\right|_{k}\equiv
\nonumber\\
\left|\left|e^{-\frac{\pi}{\Delta\lambda_q^2} \left|(r_{p,1}-r_{t,n})+
y_q^n\right|^2}\cdot \frac{\langle Q_p|Q_t\rangle + 1)}{2}\right|\right|_s \cdot 
\left|\left|e^{-\frac{\pi}{\Delta\lambda_{ \bar{q}}^2}
\left|(r_{o,1}-r_{l,n})+ y_{ \bar{q}}^n\right|^2}\cdot \frac{\langle Q_o|Q_l \rangle +1}{2}\right|\right|_k.
\label{psi}
\end{eqnarray}
As a result of the spin summation, the matrix carries subscripts
$sk$ denoting the number of quarks and antiquarks having the same
spin projections.

%----------------------------
\section{Numerical results}\label{s:results}
%----------------------------

In this section we present results of our simulations. Details of our path integral Monte Carlo simulations
have been discussed before in a variety of papers and review articles, e.g. \cite{rinton} and references therein, and will not be repeated here. The main idea of the simulations consists in constructing a Markov chain of configurations which differ by the particle coordinates (including all beads). In addition to the case of electrodynamic plasmas, here we also randomly modify the color variable $Q$ of all particles until convergence is achieved.
For the results presented below we used a cubic simulation box with periodic boundary conditions. The number of particles was equal to $N=N_q+N_{ \bar{q}}+N_g=40+40+40=120$, and the number of high-temperature factors (beads), $n=20$.

%% FIRST FIGURE: EOS
\begin {figure}[htb]
\includegraphics[width=12.cm,clip=true]{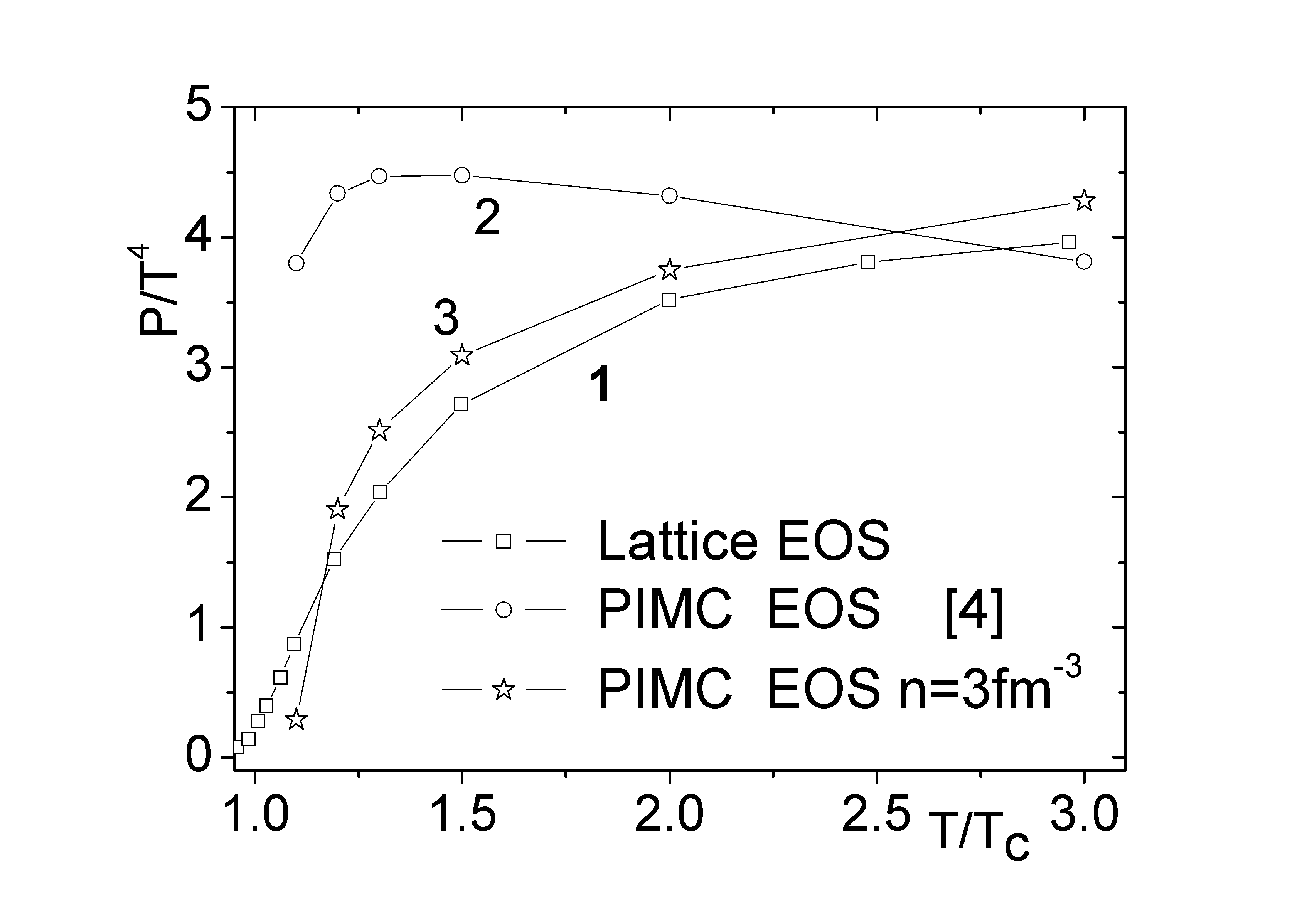}
\caption{Equation of state of the QGP from PIMC simulations compared to lattice data of Ref.~\cite{Lattice09}. Line 1 -- lattice data~\cite{Lattice09},
line 2 -- PIMC results for parametrization {\bf a},
line 3 -- PIMC results for parametrization {\bf b}, see text.}
\label{fig:EOS}
\end {figure}

In Fig.~\ref{fig:EOS} we present results for the QGP equation of state (EOS) obtained from lattice calculations, cf. Line 1,~\cite{Lattice09} and from our PIMC calculations according to Eq.~(\ref{e_gen}) 
with $N_q=N_{ \bar{q}}=N_g=N/3$, based on two different parametrizations of the quasiparticle model: 
\begin{description}
 \item[a] Temperature dependence of quasiparticle density, mass and coupling constant are chosen according to $m(T)/T_c=0.9/(T/T_c-1)+3.45+0.4T/T_c$~\cite{LiaoShuryak,shuryak1}. Results for  $n(T)=0.244 T^3$ and $g^2(T)=2\pi$ are shown by Line 2 in Fig.~\ref{fig:EOS}.
 \item[b] Same mass and coupling constant as in a, but with $T$-independent  quasiparticle density, $n\approx 3 fm^{-3}$. The results are shown by Line 3.
\end{description}
As seen from Fig.~\ref{fig:EOS}, the version with the $T$-dependent density ({\bf a}) results in
substantial deviations from the lattice data.
In contrast, the constant-density version ({\bf a}) gives a surprisingly good agreement with the lattice results in the whole range of temperatures down to values as low as $T=1.2 T_c$. Despite the simplicity of model {\bf b} 
it seems to capture basic trends of the global thermodynamic properties of the QGP. 

With the PIMC simulations we are now able to analyze more in detail additional properties and the internal structure of the QGP which can be understood from the pair distribution functions, see below. First we note 
that the QGP in the studied temperature range is, in fact, quantum degenerate. This follows from the degeneracy parameter $\chi_a = n_a\Lambda_a^3$, where $\Lambda=h/\sqrt{2\pi m_a k_BT}$ is the thermal De Broglie wave length of the quasiparticle of species ``a'' (here it is the same for quarks, and gluons). In the studied temperature interval, $\chi$ is practically constant and equal $4.1$. 
From this we expect that the finite extension of the quasiparticles is relevant and also spin statistics (e.g. the Pauli principle) should play a significant role. At the same time, the relatively moderate value of $\chi$ indicates that the chosen number of 
high-temperature factors in the PIMC simulations is appropriate. 
\begin{figure}[htb]\label{fig:cor}
\vspace{0cm} \hspace{0.0cm}
\includegraphics[width=8.cm,clip=true]{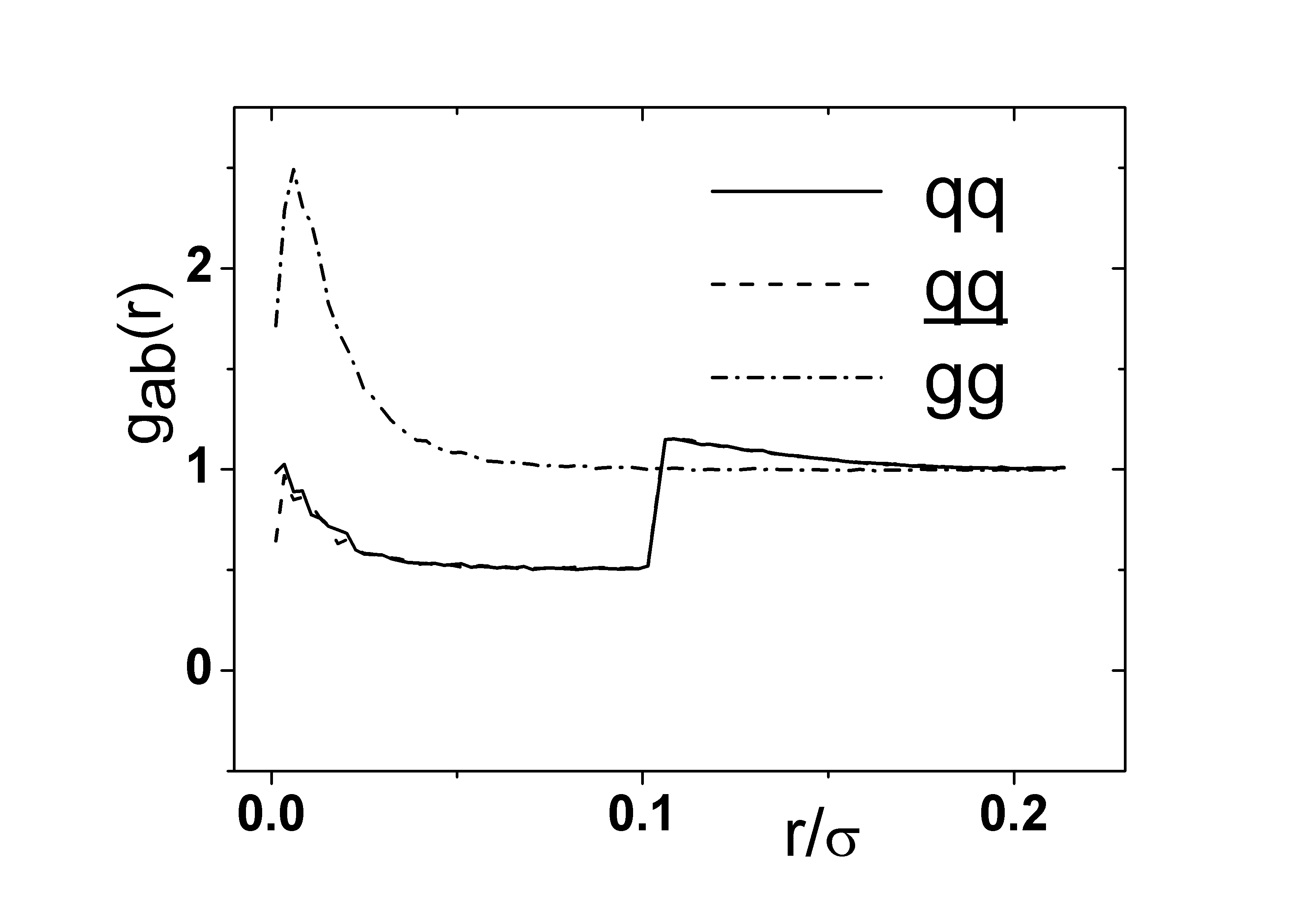}
\includegraphics[width=8.cm,clip=true]{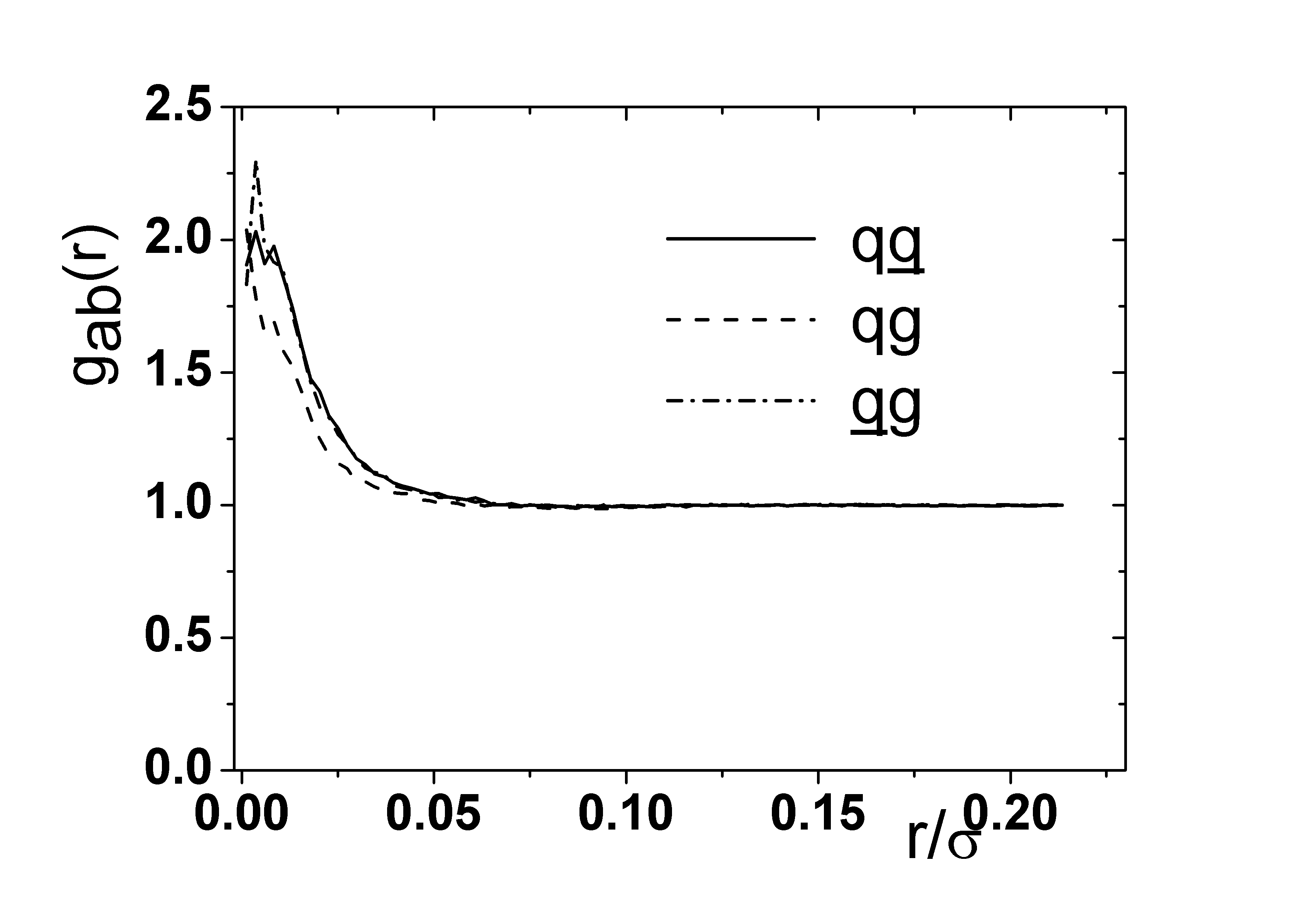}
\includegraphics[width=8.cm,clip=true]{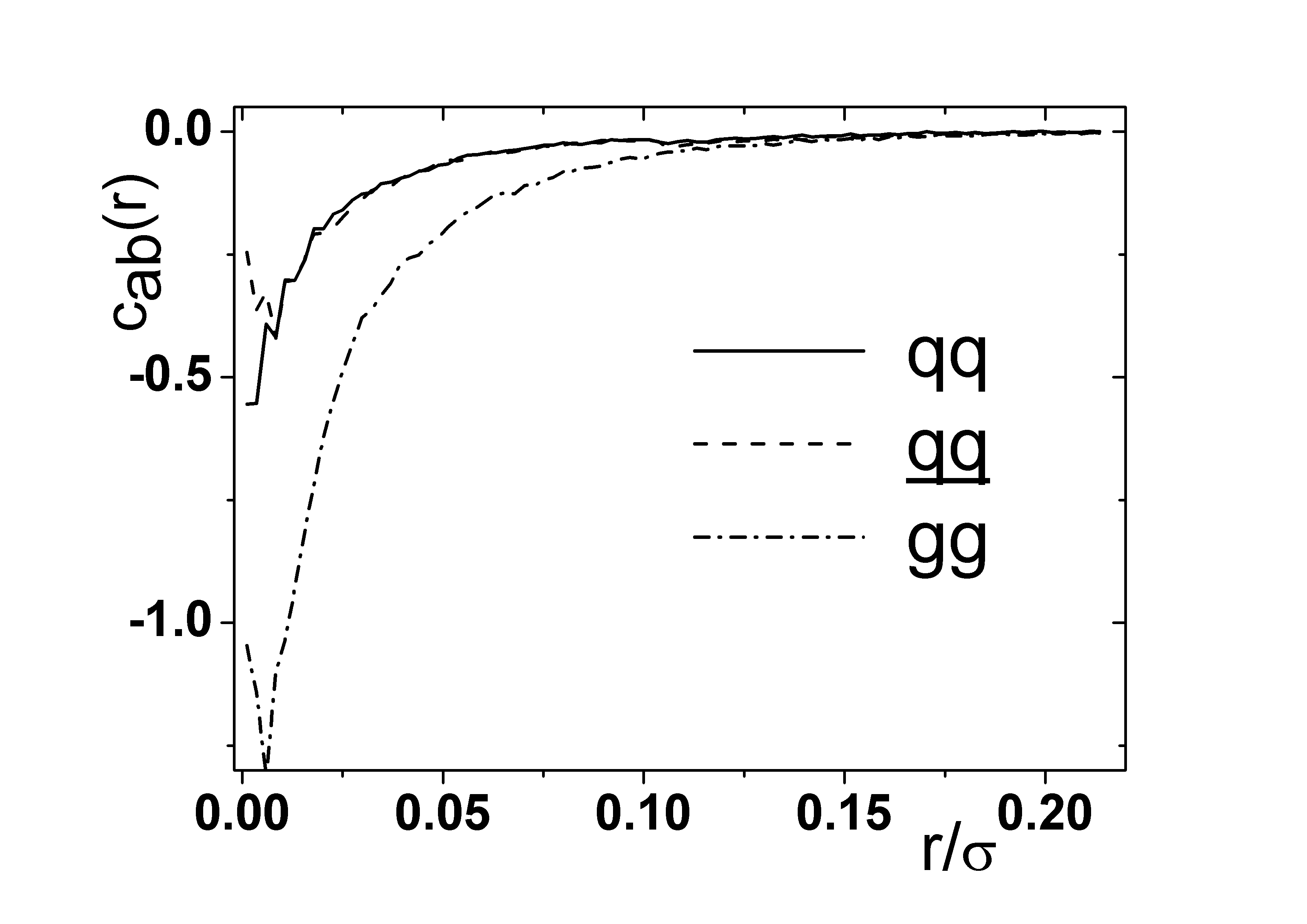}
\includegraphics[width=8.cm,clip=true]{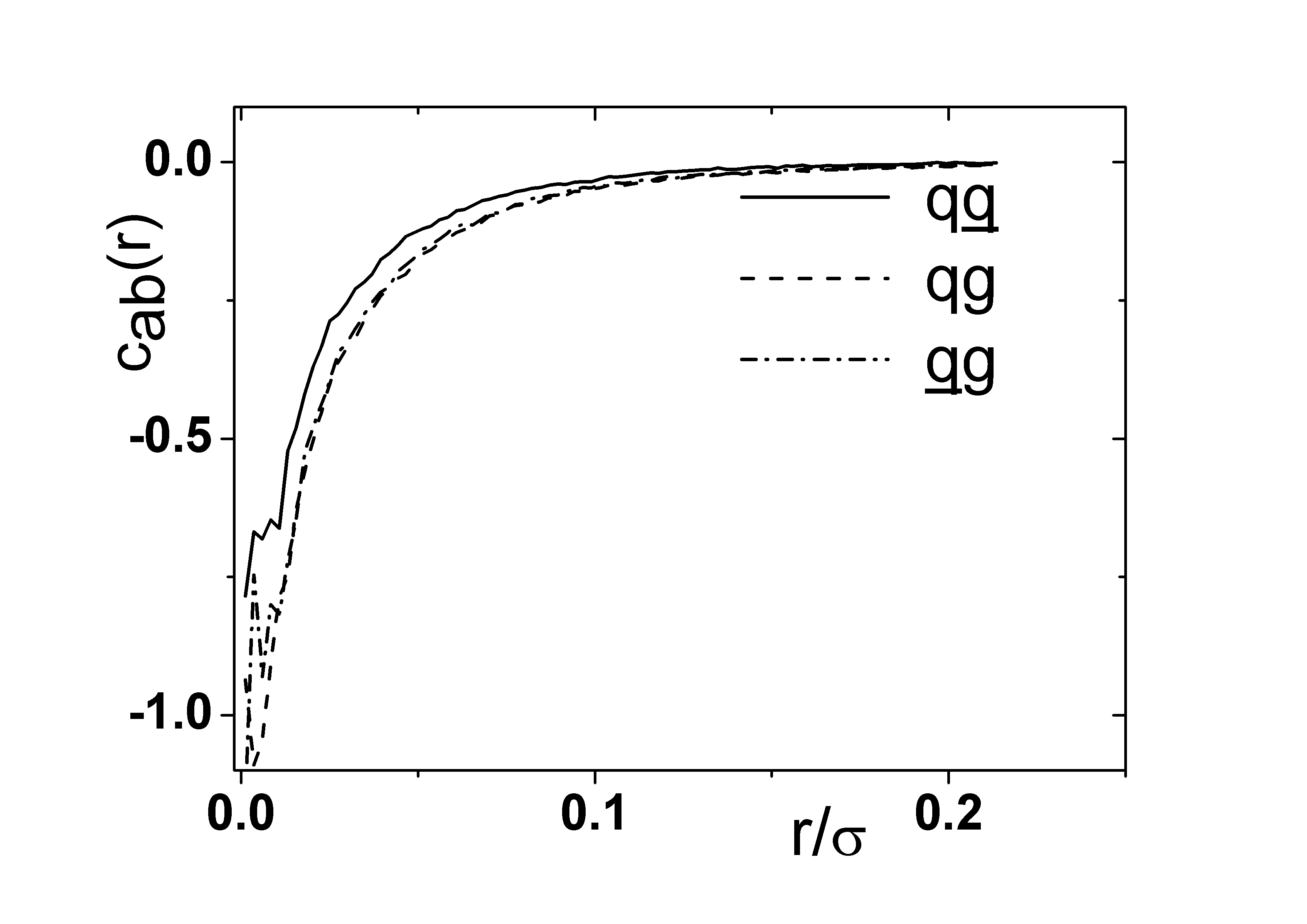}
\caption{Pair distribution functions (upper panel) and color pair distribution
functions (lower panel) of identical particles (left column) and different particles (right column) at 
temperature $T/T_c=3$ for the parametrization {\bf b}. The mean interparticle distance is 
$\langle r \rangle/\sigma=0.06$, where the length scale is defined as $\sigma=2 \pi \hbar c/k_B T_c=7.27$ fm and $T_c=175$MeV.
}
\end{figure}

Let us now consider the spatial arrangement of the quasiparticles in the QGP more in detail. To this end we analyze the pair distribution and color pair distribution functions. The pair distribution functions (PDF) $g_{ab}(r)$ give the probability to find a pair of particles of type ``a'' and ``b'' at a certain distance $r$. In a non-interacting classical system, $g_{ab}(r)\equiv 1$, whereas interactions and spin effects cause re-distribution of particles. The PDF is defined according to
%%----------------------------------------------------------------------------------------------------------------
\begin{eqnarray}\label{g-def}
g_{ab}(R_1,R_2) &=& g_{ab}(R_1-R_2)=
%\nonumber\\
\frac{1}{{\tilde Z}}
\sum_{\sigma}\int\limits_V
dr dQ\,\delta(R_1-r^a_1)\delta(R_2-r^b_2)\rho(r,Q, \sigma ;\beta),
\\
{\tilde Z} &= & Z(N_q,N_{ \bar{q}},N_g,V;\beta) N_q!N_{ \bar{q}}!N_g!,
\end{eqnarray}
and results for the PDF at temperature $T/T_c=3$ are shown in Fig.~\ref{fig:cor}, top panel. Let us first consider the PDF between identical particles, see top left figure. At large distances, $r/\sigma \ge 0.15$ where $\sigma=hc/k_BT_C=7.27$fm, all functions coincide, approaching unity, as in the ideal gas case. However, there is a drastic difference in the behavior of the PDF of  quarks and gluons (the anti-quark PDF is identical to the quark PDF) and small distances. While the gluon PDF increase monotonically when the distance goes to zero, the PDF of quarks (and antiquarks) exhibits a broad minimum. This difference can be understood by spin statistics effects. In fact, for the present conditions, the De Broglie wavelength $\Lambda$ equals approximately $0.1\sigma$, i.e. the different behavior of gluons and quarks starts to appear at distances smaller than $1.5\Lambda$. The enhanced population of low distance states of gluons is due to bosonic exchange and color Coulomb attraction, see below. In contrast, the depletion of small distance states of quarks is a consequence of the Pauli principle. In an ideal Fermi gas $g(r)$ equals zero for particles with the same spin projection and one for particles with opposite spin projection, in the limit $r\to 0$. As a consequence, the spin averaged PDF approaches $0.5$ \cite{fil_kremp_bo}. This low-distance behavior is also observed in a nonideal dense astrophysical electron-ion plasma and in nonideal a electron-hole plasmas in semiconductors \cite{fil_kremp_bo,ELMP}. This is exactly the value we observe for the quark-quark PDF at small distances which we, therefore, attribute to the Pauli principle. An exception is the behavior at very small distances, $r\le 0.02 \sigma$. Also, the abrupt increase of $g$ around $r=0.1 \sigma$ is a-typical compared to electrodynamic plasmas. We, therefore expect that this behavior is caused by the particular properties of the color Coulomb interaction.

Let us now consider the PDF of different particles, see top right part of Fig.~2. %\ref{fig:cor}. 
Here all curves show similar behavior. At small distances, $r\le 0.05\sigma$, a strong increase is observed which resembles the behavior of the gluon-gluon PDF, cf. top left figure. At larger distances, all PDF's equal one. This increase of the PDF at small distances is a clear manifestation of an effective pair attraction of quarks and antiquarks as well as quarks (antiquarks) and gluons. This may, at first sight, seem surprising because all Casimir indices $C_{pt}$ are positive, see above, indicating a repulsive character of the pair potential (\ref{Coulomb}). However, this potential still contains the scalar product of the color vectors, and the net attraction could be understood if, on average, the color vectors of nearest neighbor quasiparticles of any type are anti-parallel.

This hypothesis is readily verified from the PIMC simulation data. To this end, we define the {\em color pair distribution function} (CPDF) by generalizing the definition (\ref{g-def}) according to
\begin{eqnarray}\label{c-def}
c_{ab}(R_1-R_2)=
\frac{1}{{\tilde Z}}
\sum_{\sigma}\int\limits_V
dr dQ\,\langle Q^a_1|Q^b_2 \rangle \delta(R_1-r^a_1)\delta(R_2-r^b_2)\rho(r,Q, \sigma ;\beta),
\end{eqnarray}
which is straightforwardly computed during the PIMC simulations together with the traditional PDF. The results 
are shown in the lower panel of Fig.~2. %\ref{fig:cor}. 
We immediately observe that all CPDF's are negative at small distances indicating anti-parallel orientation of the color vectors of all neighboring quarks (antiquarks) and gluons as well as quarks and antiquarks, clearly confirming the origin of the effective quasiparticle attraction seen in the functions $g_{ab}$ for $a\ne b$. 
We now turn to the CPDF of identical particles, see bottom left figure. All functions are non-positive everywhere. 
The minimum of $c_{qq}$ close to $r=0$ explains the increase of $g_{qq}$ above the value $0.5$ at small distances. Most striking is the deep minimum of the gluon CPDF, $c_{gg}$, at small distances. It again confirms the antiparallel arrangement of the color vectors of neighboring gluons whereas the much lower value of the minimum, compared to that of the quark CPDF, is due to the absence of the Pauli principle and the larger value of the Casimir index $C_{gg}$ compared to $C_{qq}$. This deep minimum explains the high maximum of the gluon PDF $g_{gg}$.

Let us summarize the local ordering of the QGP at the temperature $T/T_c=3$. We observe only weak signs of a spatial ordering, cf. the peak of the quark PDF around $r=0.11\sigma$, which may be interpreted as emergence of liquid-like behavior of the QGP. Much more pronounced is the short range structure of nearest neighbors. The QGP lowers its total energy by minimizing the color Coulomb interaction energy via a spontaneous ``anti-ferromagnetic'' ordering of color vectors of gluons. 
This gives rise to a clustering of gluons which is accompanied by a weak tendency of clustering of quark pairs with anti-parallel spins. We also observe 
 clusters of quarks, antiquarks and gluons. To verify the relevance of these trends a more refined spin-resolved analysis of the PDF and CPDF is necessary, together with simulations
in a broader range of temperatures which are presently under way.

%----------------------------
\section{Discussion}\label{s:discussion}

Experimental data on the quark-gluon plasma and the hadronization transition give rise
to numerous challenges to the theory, see, e.g. \cite{shuryak08,biro08} and references therein.
Of particular interest is the question why the
quark-gluon matter behaves as an almost perfect fluid rather than as a perfect gas, as it
could be expected from the asymptotic freedom.
Quantum Monte Carlo simulations based on the quasiparticle picture with color Coulomb interactions help us to answer this question. Indeed, the ratio of the potential energy
of the system to the kinetic one, obtained in these simulations, turns out to be in the range from $1$ to $3$, depending on the temperature. This certainly corresponds to a liquid-like rather than a gas-like  behavior.

We have shown that the PIMC method captures main trends of the equation of state (even near the critical temperature) and may also yield valuable insight into the internal structure of the QGP, in particular into the pair correlation functions.
Our PIMC simulations also allow for a selfconsistent analysis of cluster and bound state formation in the QGP.
Similar questions have been successfully studied before
in dense astrophysical plasmas \cite{filinov_jetpl00} and electron-hole plasmas in semiconductors \cite{filinov_jpa03}. In fact, first indications for clustering in the QGP have been observed and will be studied in more detail in the future.

The PIMC method is not able to yield dynamical and transport properties of the QGP. One way to achieve this is to develop semiclassical molecular dynamics simulations. In contrast to previous MD simulations where quantum effects were included phenomenologically via a short range potential \cite{shuryak1} a more systematic approach has been developed for electron-ion plasmas \cite{afilinov_jpa03,afilinov_pre04}. There an effective quantum pair potential has been derived from quantum Monte Carlo data which should also be possible in application to the QGP. Finally, another very promising approach to study the dynamical and transport properties of strongly coupled Coulomb systems is based on the Wigner formulation of quantum dynamics \cite{FilThom} which should also be applicable to the  quark gluon plasma.

%----------------------------
\section*{Acknowledgements}
We acknowledge stimulating discussions with Prof.~B.~Friman and Prof.~M.I.~Polikarpov and financial support by the \emph{Innovationsfond  Schleswig-Holstein}. Y.I. and V.S. were partially supported by the Bundesministerium f\"ur Bildung und Forschung (BMBF project RUS 08/038). Y.I. acknowledges support of the
Russian Federal Agency for Science and Innovations (grant NSh-3004.2008.2).

%----------------------------

\bibliographystyle {apsrev}

\end{document}